\documentclass[conference]{IEEEtran}
\IEEEoverridecommandlockouts

\usepackage{float}
\usepackage{bm}
\usepackage{cite}
\usepackage{amsmath,amssymb,amsfonts}
\usepackage{algorithm}
\usepackage{algorithmic}
\usepackage{graphicx}
\usepackage{textcomp}
\usepackage{xcolor}
\usepackage{epstopdf}
\usepackage{subfigure}
\usepackage{authblk}
\usepackage{xcolor}
\usepackage{setspace}

\newcommand{\Rmnum}[1]{\uppercase\expandafter{\romannumeral #1\relax}}

\def\BibTeX{{\rm B\kern-.05em{\sc i\kern-.025em b}\kern-.08em
    T\kern-.1667em\lower.7ex\hbox{E}\kern-.125emX}}
\begin{document}
\title{Dynamic Content Update for Wireless Edge Caching via Deep Reinforcement Learning}
\author{Pingyang Wu, Jun Li, Long Shi, Ming Ding, Kui Cai, and Fuli Yang
\thanks{Manuscript received xxx xxx, xxx; revised xxx xxx, xxx; accepted xxx xxx, xxx. The work of P. Wu and J. Li was supported   by National Key R\&D Program under Grant 2018YFB1004800 and National Natural Science Foundation of China under Grants 61872184 and 61727802. The work of L. Shi and K. Cai was supported by Singapore Ministry of Education Academic Research Fund Tier 2 MOE2016-T2-2-054 and SUTD-ZJU Grant ZJURP1500102. \emph{(Corresponding authors: Long Shi, Jun Li.)}}
\thanks{P. Wu and J. Li are with the School of Electronic and Optical Engineering, Nanjing University of Science and Technology, Nanjing 210094, China (e-mail: \{pingyang.wu, jun.li\}@njust.edu.cn). L. Shi and K. Cai are with the Science and Math Cluster, Singapore University of Technology and Design, Singapore 487372 (e-mail: slong1007@gmail.com, cai\_kui@sutd.edu.sg). M. Ding is with the Data61, CSIRO, Sydney, N.S.W. 2015, Australia (e-mail: Ming.Ding@data61.csiro.au). F. Yang is with the China Unicom Jiangsu Branch, Nanjing 210019, China (e-mail: fuliyang@chinaunicom.cn).}}
\maketitle
\begin{abstract}
This letter studies a basic wireless caching network where a source server is connected to a cache-enabled base station (BS) that serves multiple requesting users. A critical problem is how to improve cache hit rate under dynamic content popularity. To solve this problem, the primary contribution of this work is to develop a novel dynamic content update strategy with the aid of deep reinforcement learning. Considering that the BS is unaware of content popularities, the proposed strategy dynamically updates the BS cache according to the time-varying requests and the BS cached contents. Towards this end, we model the problem of cache update as a Markov decision process and put forth an efficient algorithm that builds upon the long short-term memory network and external memory to enhance the decision making ability of the BS. Simulation results show that the proposed algorithm can achieve not only a higher average reward than deep Q-network, but also a higher cache hit rate than the existing replacement policies such as the least recently used, first-in first-out, and deep Q-network based algorithms.
\end{abstract}
\IEEEpeerreviewmaketitle
\section{Introduction}
The rapid increase in the number of ubiquitous wireless devices will inevitably produce the sheer volume of traffic load, resulting in the network congestion in the near future. With the advent of the 5G networks, caching at the wireless edge has been used to accelerate the content download speed and improve the performance of wireless networks~\cite{1}. Wireless caching features high temporal variability of the user requests. To meet the time-varying requests, base stations (BSs) with limited cache size frequently replace their local caches according to cache replacement policies, e.g., the least recently used (LRU) and first-in first-out (FIFO)~\cite{lru2, fifo2}.

Due to the complexity of the real environment, these conventional replacement policies cannot accurately capture dynamic characteristics of content popularity~\cite{pop}. Inspired by the reinforcement learning (RL) in solving complicated control problem~\cite{RL}, the works in~\cite{drl},~\cite{new1} relied on strong feature representation ability of deep neural network (DNN)~\cite{new2} and adopted the model-free deep RL (DRL) to maximize the long-term system reward in mobile edge caching. In~\cite{drl,new1,new2}, the edge node fetches the missed content from the source server and replaces its local cache with newly fetched content. However, it is possible that the newly fetched content is less popular than the cached content. In this context, the fetch-and-replace strategy in the cache miss case may not be efficient.

Driven by this issue, we propose a novel content update strategy in the wireless caching network to improve cache hit rate in the BS. To our best knowledge, few existing work on the cache replacement has taken into account either dynamic characteristics of content popularity~\cite{pop} or advanced content update strategy rather than the intuitive fetch-and-replace strategy in~\cite{drl},~\cite{new2}. The update strategy evicts or retains content in the BS by taking both the BS cache and user requests into consideration (see Section \ref{sec:update}). We first formulate the problem of content update as a Markov decision process (MDP) with the state space consisting of the BS cache and user requests and the action space including evicting and retaining (see Section \ref{sec:mdp}). Then, we put forth a DRL-based algorithm to enhance the decision making ability of the BS, by leveraging the long short-term memory (LSTM) network and external memory (see Section \ref{sec:alg}). Our simulation results show that, superior to LRU and FIFO replacement and deep Q-network (DQN) algorithm, the proposed external memory-based recurrent Q-network (EMRQN) algorithm significantly improves cache hit rate and long-term system reward.
\section{System Model}\label{sec:sys}
Considers a basic wireless caching network consisting of a source server, a single cache-enabled BS, and $K$ users, where the BS is connected to the source server through wireless backhaul. Let $\mathcal{O}=\lbrace o_1,o_2, \ldots, o_{|\mathcal{O}|} \rbrace$ denote a set that collects all $|\mathcal{O}|$ contents in the server. The BS with limited cache storage can predownload $N$ contents from the server. Suppose that contents in the server cover all possible requests from all users in real time.

This letter studies a caching scenario where only a small portion of contents in the server are requested and thereby prefetched by the BS. That is, $|\mathcal{O}| \gg N$. Given the limited cache storage of the BS, the maximum number of contents requested by each user is $N$. Consider that the BS can receive these requests from multiple users without knowing content popularities. To efficiently meet the time-varying requests, the BS should update its local cache accordingly.
\section{Proposed Cache Update Strategy}\label{sec:update}
This section first shows a flowchart of content delivery and content update and then illustrates the content update procedure by a toy example.
\subsection{Flowchart}
\begin{figure}[t]
\setlength{\abovecaptionskip}{0.cm}
\setlength{\belowcaptionskip}{-0.cm}
\centerline{\includegraphics[width=3.6in]{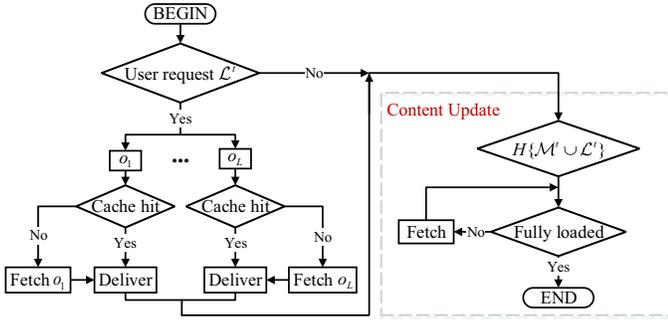}}
\caption{The flowchart of content delivery and content update.}
\label{fig:flow}
\end{figure}

Fig. \ref{fig:flow} shows the flowchart of the caching strategy consisting of conventional delivery phase and proposed content update phase. This system operates in a discrete time fashion with time slot $t\in \mathcal{T}=\{1,2,\ldots,T\}$ and integer $T \leq \infty$. Let $\mathcal{M}^t$ denote a set of contents cached in the BS in time slot $t$. Consider that the BS cache is fully loaded in any time slot (i.e., $|\mathcal{M}^t|=N$). The distinct contents requested by $K$ users are included in $\mathcal{L}^t=\{o_1, o_2, \ldots, o_L\}$.

In the delivery phase, if the requested content $o_n$ is stored in the BS (i.e., cache hit), the BS directly delivers $o_n$ to the user. Otherwise, if $o_n$ is missed in the BS cache (i.e., cache miss), the BS fetches $o_n$ from the server and delivers it to the user. As such, all user requests are fulfilled.

Existing works in~\cite{lru2,pop,drl} directly replace the BS cache with the newly fetched contents. In the content update phase, we propose to \emph{update} the BS cache by taking into account both the newly fetched contents and its cache in current time slot. The BS first \emph{evicts} or \emph{retains} some contents in $\mathcal{M}^t\cup \mathcal{L}^t$ according to an action indicator $H\{\cdot\}$. As Section \ref{sub:as} will elaborate, the BS evicts content $o$ if $H\{o\}=0$ or retains $o$ if $H\{o\}=1$. Second, the BS checks if the current cache is fully loaded after evicting or retaining. If the BS cache is fully loaded, the procedure ceases. Otherwise, the BS fetches new contents with high normalized \emph{cumulative request} from the server to fully load the BS cache (see Section \ref{sub:ss}).

As Section \ref{sec:alg} will elaborate, the BS updates its cache by a DRL-based algorithm. It is known that the decision making in the DRL is not perfect. Therefore, the BS still needs to update its cache based on $\mathcal{M}^t$ only to further improve the update accuracy, even if there is no user request (i.e., $\mathcal{L}^t=\varnothing$).
\subsection{A Toy Example}
\begin{figure}[t]
\setlength{\abovecaptionskip}{0.cm}
\setlength{\belowcaptionskip}{-0.cm}
\centerline{\includegraphics[width=2.8in,angle=0]{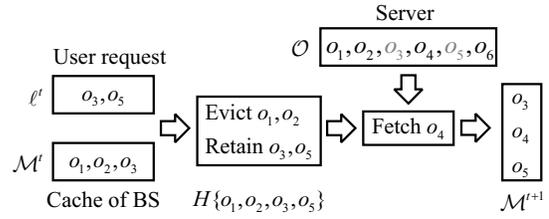}}
\caption{An illustrating example of proposed content update.}
\label{fig:toy}
\end{figure}

Let us see a toy example in Fig. \ref{fig:toy} to illustrate the proposed content update. Consider that the server owns $\mathcal{O}=\{o_1,o_2,o_3,o_4,o_5,o_6\}$, the BS is fully loaded in time slot $t$ by storing $\mathcal{M}^t=\{o_1, o_2, o_3\}$, and users request $\mathcal{L}^t=\{o_3, o_5\}$. First, the BS fetches $o_5$ from the server and delivers $\{o_3, o_5\}$ to the requesting users. Second, according to $H\{o_1, o_2, o_3, o_5\}=\{0, 0, 1, 1\}$, the BS evicts $o_1,o_2$ and retains $o_3,o_5$.
To be fully loaded, the BS fetches $o_4$ from the server, as $o_4$ has the largest normalized cumulative request among $\{o_1, o_2, o_4, o_6\}$. Finally, the BS cache in time slot $t+1$ is updated as $\{o_3, o_4,  o_5\}$.
\section{MDP Formulation}\label{sec:mdp}
\subsection{State Space}\label{sub:ss}
Without loss of generality, consider that the BS stores $\mathcal{M}^t=\lbrace o_1, o_2,\ldots,o_N \rbrace$ in time slot $t$. Let $\mathcal{S}=\{s|s=\langle \mathcal{B},\ \mathcal{L} \rangle \}$ be the set of system state space. In this set, each system state $s^t$, consisting of a BS state $\mathcal{B}^t$ and a user request state $\mathcal{L}^t$ in time slot $t$, is given by $s^t=\langle \mathcal{B}^t,\ \mathcal{L}^t \rangle$, where $\mathcal{L}^t = \{\ell^t_1, \ell^t_2, \ldots, \ell^t_K\}$ with $\ell^t_k$ being a set that collects the contents requested by user $k$ in time slot $t$. We stress that $\ell^t_k$ can be any subset of $\mathcal{O}$ with $|\ell^t_k| \leq N, \forall k$, due to the limited cache size in the BS. If $\ell^t_k = \varnothing$, there is no request from user $k$ in time slot $t$. Moreover, $\mathcal{B}^t = \{(o_n,\ q_{o_n}^t)|o_n \in \mathcal{M}^t \}$, includes content $o_n$ in the BS cache and the normalized cumulative request $q^t_{o_n}$ of $o_n$ in time slot $t$, which evolves in a time-homogeneous Markov chain as $q_{o_n}^{t+1} = q_{o_n}^{t} + \delta^t_{o_n},\ n \in \lbrace 1,2,\ldots,N \rbrace$, where $\delta^t_{o_n} = \frac{c_{o_n}^{t}}{\sum_{n'=1}^{N}c_{o_{n'}}^{t}}$ denotes the normalized number of request times of $o_n$ in time slot $t$ with $c_{o_n}^t$ being the times of $o_n$ requested in time slot $t$. In addition, $q_{o_n}^{t+1}$ accumulates the number of times that $o_n$ is requested over time slots $\{1,2,\ldots,t\}$ with increment of $\delta_{o_n}^t$.

If there is no user request, the BS updates its cache based on $\mathcal{B}^t$ only. In this case, the BS will evict $o_n$ if its associated $q_{o_n}^t$ is below a designated threshold $\bar{q}^t = \frac{\sum_{n=1}^{N}q^t_{o_n}}{N}$~\cite{th}.
\subsection{Action Space}\label{sub:as}
Given any content $o$, the BS decides whether to evict or retain this content by a binary indicator $H\{o\} \in \{0, 1\}$. If $H\{o\} = 0$, the BS evicts $o$, otherwise the BS retains $o$. Given any system state $s^t$, the BS carries out action in time slot $t$ according to
\begin{equation}
    \alpha^t_s=
\begin{cases}
    H\{\mathcal{M}^t\},&\rm{if}\ \mathcal{L}^t \subseteq \mathcal{M}^t\ \rm{or}\ \mathcal{L}^t=\varnothing \\[0.5ex]
    H\{\mathcal{M}^t \cup \mathcal{L}^t\},&\rm{if}\ \! \mathcal{M}^t \cap \mathcal{L}^t \! \neq \! \varnothing\ \! \rm{and}\ \! \mathcal{L}^t \! \nsubseteq \! \mathcal{M}^t
\end{cases},\tag{1}
\end{equation}
where $\alpha^t_s$ is a collection that contains 0's or 1's. To be specific, the BS only updates its own cache $\mathcal{M}^t$, if all user requests hit (i.e., $\mathcal{L}^t \subseteq \mathcal{M}^t$) or there is no user request (i.e., $\mathcal{L}^t=\varnothing$). Otherwise, the BS updates contents $\mathcal{M}^t \cup \mathcal{L}^t$ consisting of its own cache and newly fetched contents, if some user requests miss (i.e., $\mathcal{M}^t \cap \mathcal{L}^t \neq \varnothing \ \rm{and} \ \mathcal{L}^t \nsubseteq \mathcal{M}^t$).

Consequently, the action space corresponding to the state space $\mathcal{S}$ can be expressed as $\mathcal{A} = \bigcup_{t\in \mathcal{T}}\alpha^t_s,\ \forall s\in \mathcal{S}$.
\begin{algorithm}[t]
\caption{EMRQN for Dynamic Content Update}
\label{alg:EMRQN}
\begin{algorithmic}[1]
\setstretch{0.9}
\REQUIRE ~~\\
    \STATE $Q$-value and network parameter; cache size of the BS; the long-term reward $G = 0$; the average reward $g = 0$.
\ENSURE ~~\\
    \FOR{episode = 1 to $E$}
    \STATE Initialize system state $s^0$;\\
    \FOR{$t$ = 1 to $T$}
    \STATE Update parameters according to $\varepsilon$-greedy method~\cite{RL};\\
    \STATE Select the action $\alpha_s^t \!=\! \arg\max Q(s^t,\alpha_s^t)$ with probability of $1\!-\! \varepsilon$; or randomly select an action with probability of $\varepsilon$;\\
    \STATE Take \hspace{-0.3mm}action \hspace{-0.3mm}$\alpha_s^t$, \hspace{-0.3mm}receive \hspace{-0.3mm}a \hspace{-0.3mm}reward \hspace{-0.3mm}$R^t$ \hspace{-0.3mm}and \hspace{-0.3mm}next \hspace{-0.3mm}state $s^{t+1}$;\\
    \STATE Store \hspace{-0.3mm}transition \hspace{-0.3mm}$(s^t,\alpha_s^t,R^t,s^{t+1})$ \hspace{-0.3mm}in \hspace{-0.3mm}experience \hspace{-0.3mm}replay;\\
    \STATE Compute states similarity and modify $Q$-value for state-action pairs according to \eqref{eq:modify};\\
    \STATE Update $Q$-value and network parameter;\\
    \STATE Calculate $G^t$.
    \ENDFOR
    \STATE $g = \frac{\sum_{t=1}^{T}G^t}{T}$.
    \ENDFOR
\end{algorithmic}
\end{algorithm}
\subsection{Reward Function}\label{sub:reward}
Let $\mathcal{D}^t_+$ denote the set that collects the newly cached contents in the BS in time slot $t$, and $\mathcal{D}^t_*$ denote the set that collects the contents not only cached in time slot $t-1$ but also retained in time slot $t$. In this context, we design the \emph{positive} reward as
\begin{equation}
    R^{t}_{+}(s^t,\alpha^t_s)=\sum_{o_{*} \in \mathcal{D}^t_*} v(c_{o_{*}}^t) + \sum_{o_{+} \in \mathcal{D}^t_+} \eta v(c_{o_{+}}^t), \tag{2}
\end{equation}
where $v(c_{o}^t)$ is the normalized amount of requests for content $o \in \mathcal{D}^t_{*} \cup \mathcal{D}^t_{+}$, to represent the reward induced by content delivery in time slot $t$. Note that the fetching of $o_{+} \in \mathcal{D}^t_+$ from the server in the cache miss case deserves a scaled reward by $0<\eta<1$, where $\eta$ is used to bias the BS toward improving cache hit rate.

In addition, let $\mathcal{D}^t_-$ be the set that collects the contents evicted in time slot $t-1$. In some cases, we find that the cache miss occurs in time slot $t$, when the content is evicted in time slot $t-1$ but is requested in time slot $t$. In view of this, we define a \emph{negative} reward as
\begin{equation}
    R^{t}_{-}(s^t,\alpha^t_s)=\sum_{o \in \mathcal{D}^t_+ \cup \mathcal{D}^t_-} m(c_{o}^t)\ , \tag{3}
\end{equation}
where $m(c_{o}^t)$ is the normalized amount of requests for content $o \in \mathcal{D}^t_{+} \cup \mathcal{D}^t_{-}$, to represent the cost caused by evicting or fetching in time slot $t$.

Finally, the immediate system reward induced by action $\alpha^t_s$ at state $s^t$ in time slot $t$ is given by~\cite{reward}
\begin{equation}
    R^{t}(s^t,\alpha^t_s) = R^{t}_{+}(s^t,\alpha^t_s)-R^{t}_{-}(s^t,\alpha^t_s), \tag{4} \label{eq:reward}
\end{equation}
which is used to reward the BS with the cache hit and punish the BS for the cache miss.
\section{External Memory-based Recurrent Q-Network}\label{sec:alg}
\begin{figure}[t]
\setlength{\abovecaptionskip}{0.cm}
\setlength{\belowcaptionskip}{-0.cm}
\centerline{\includegraphics[width=3.2in]{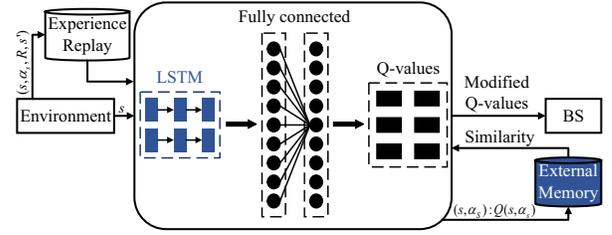}}
\caption{The architecture of the proposed EMRQN algorithm.}
\label{fig:alg}
\end{figure}

In this section, we propose the EMRQN algorithm (see Algorithm \ref{alg:EMRQN}) to maximize the long-term system reward $G^t=\sum_{k=0}^{\infty}\gamma^{k}R^{t+k}$, where $R^t$ is defined in \eqref{eq:reward} and the discount factor $\gamma$ ranges between 0 and 1~\cite{RL}. The output of Algorithm \ref{alg:EMRQN} is the average reward $g=\frac{\sum_{t=1}^{T}G^t}{T}$ (see step 13 of Algorithm \ref{alg:EMRQN}). Building upon DQN, we tentatively employs LSTM to enable the BS with stronger decision making ability as well as external memory to modify the $Q$-value as shown in Fig. \ref{fig:alg}. Note that this work adopts the DNN for function approximation, since DNN has better characterization and generalization ability than linear function approximation~\cite{RL}.
\subsection{Long Short-Term Memory Recurrent Network}
LSTM can alleviate the vanishing gradient problem of common NN and recurrent neural network (RNN), which provides an easy path for gradient flow during back-propagation~\cite{dl}. In the sequential decision-making problem, LSTM can extract useful information from historical data and incorporate contextual information from past inputs to predict $Q(s,\alpha_s)$ of the current state-action pair $(s,\alpha_s)$. We first determine the value of $H\{\cdot\}$ from step 6 of Algorithm \ref{alg:EMRQN} for each time slot, and then the BS decides to either evict or retain content based on this $H\{\cdot\}$. With the aid of LSTM, the BS can make better decisions by using historical data effectively.
\subsection{External Memory}
We use a finite-size external memory to store $(s,\alpha_s)$ and the corresponding maximum $Q$-value. Note that the external memory discards the first stored samples if it is full. Let $s_{ex}=\{ \mathcal{B}_{ex},\mathcal{L}_{ex}\}$ denote the system state in external memory, where $\mathcal{B}_{ex}$ and $\mathcal{L}_{ex}$ represent the BS state and user requests in external memory respectively, and $\mathcal{M}_{ex}$ denotes cached contents from $\mathcal{B}_{ex}$. In order to improve prediction model accuracy, we follow the neighborhood method in~\cite{nn} to modify the $Q$-value, where the BS takes similar actions in the like-minded states. First, the similarity between $s$ and $s_{ex}$ is given by $\mathrm{sim}(s, s_{ex})=1/\big(1+\mathrm{d}(s, s_{ex})\big)$, where
\begin{align}
    \mathrm{d}(s, s_{ex})= \big(&\sum\limits_{o_i\in \mathcal{M}\cap \mathcal{M}_{ex}}(H\{o_i\}-H_{ex}\{o_i\})^2+  \nonumber \\
                            & \sum\limits_{o_j\in \mathcal{L}\cap \mathcal{L}_{ex}}(H\{o_j\}-H_{ex}\{o_j\})^2\big)^{\frac{1}{2}},  \tag{5}
\end{align}
denotes the Euclidean distance between $s$ and $s_{ex}$ with $H_{ex}\{o\}$ being the action indicator of $o$ in external memory. Then the $Q$-value is modified as
\begin{align}
Q_{re}(s,\alpha_s&)=Q(s,\alpha_s)+\nonumber \\
                         &\frac{\sum\limits_{s_{ex}\in \mathcal{S}_{ex}}\! \mathrm{sim}(s,s_{ex})[Q(s_{ex},\alpha_s)\!-\!Q(s,\alpha_s)]}{\sum\limits_{s_{ex} \in \mathcal{S}_{ex}}\! |\mathrm{sim}(s, s_{ex})|}, \tag{6} \label{eq:modify}
\end{align}
where $\mathcal{S}_{ex}$ is the set that collects all possible system states in external memory.
\begin{figure}[t]
\setlength{\abovecaptionskip}{0.cm}
\setlength{\belowcaptionskip}{-0.cm}
\centerline{\includegraphics[width=3.2in,angle=0]{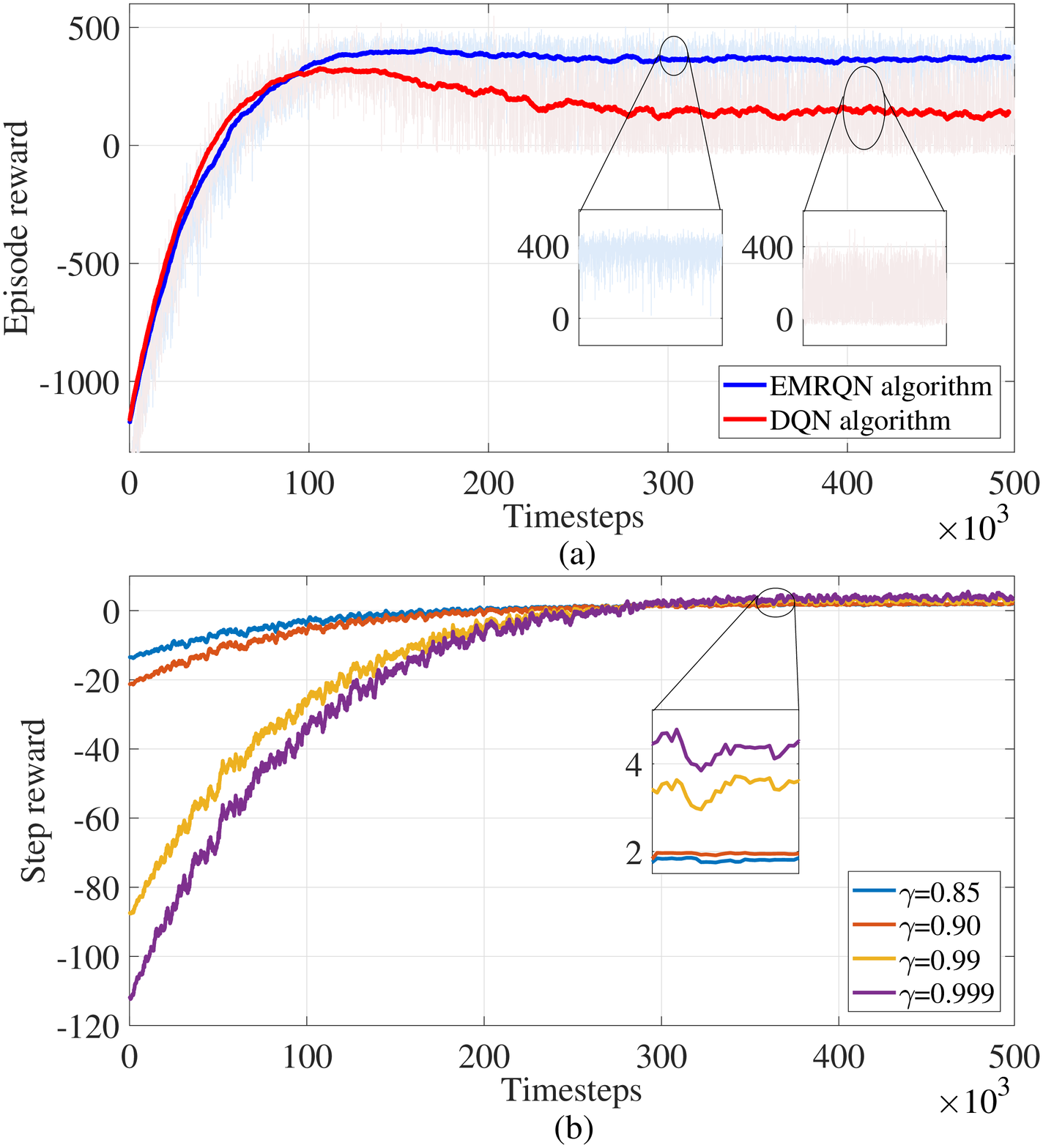}}
\caption{(a) Comparison of average episode reward between EMRQN and DQN. (b) Comparison of average step reward under different $\gamma$.}
\label{fig:reward}
\end{figure}
\begin{figure}[t]
\setlength{\abovecaptionskip}{0.cm}
\setlength{\belowcaptionskip}{-0.cm}
\centerline{\includegraphics[width=3.2in,angle=0]{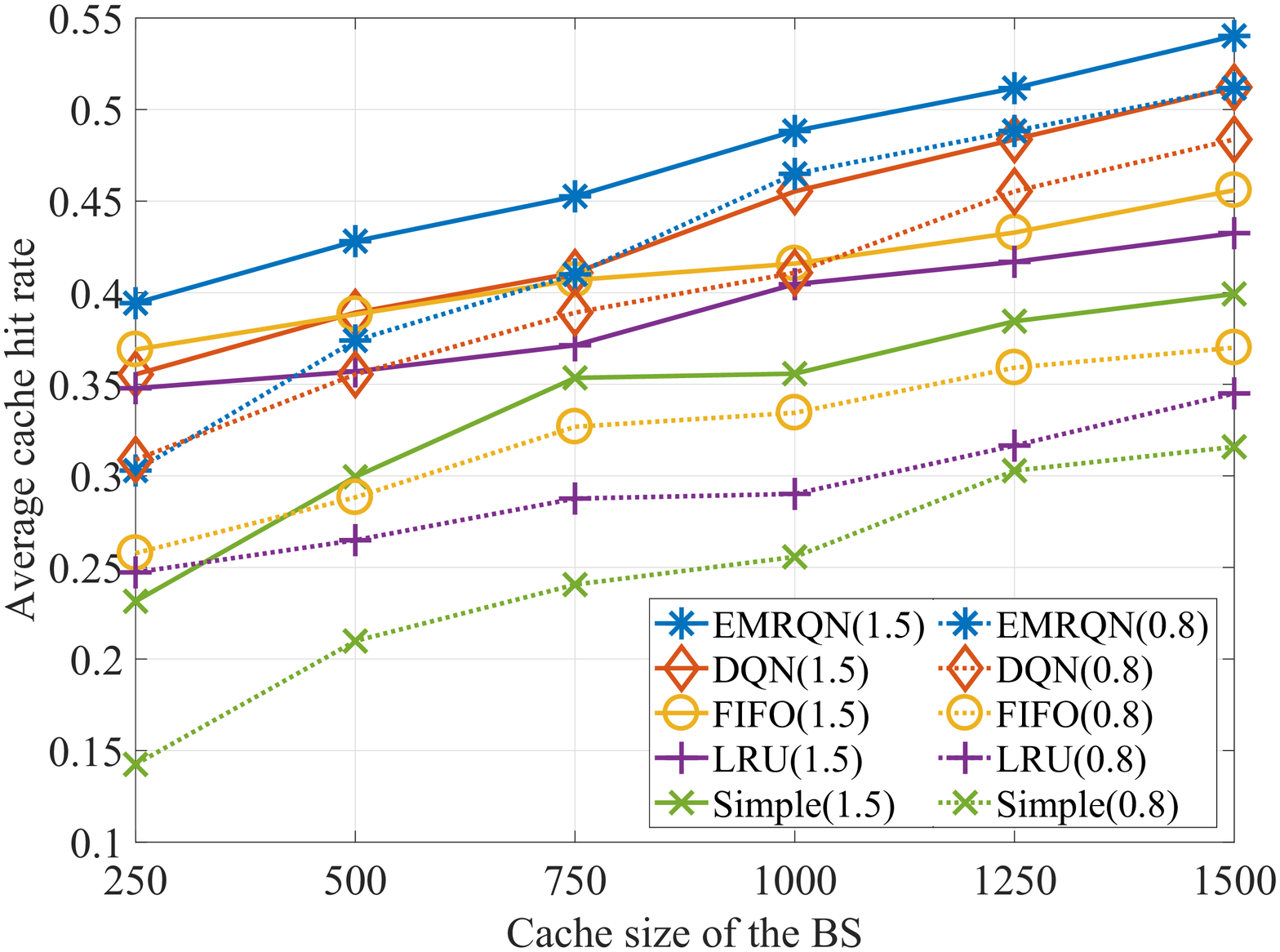}}
\caption{Comparison of average cache hit rate under different cache sizes.}
\label{fig:hit}
\end{figure}
\section{Simulation Results}\label{sec:simu}
In this section, we compare the performance of proposed EMRQN algorithm with LRU, FIFO, and DQN algorithms~\cite{drl}. Note that LRU always evicts the least recently used content, FIFO evicts the first cached content. In the simulation results, we initialize the probability of choosing a random action to be 1 and decay exponentially towards 0.01. We use an external memory size of 80000 and PyTorch as DNN framework, where Adam optimizer chooses weight decay of 0.00001 and batch size of 8 to adjust the effect of model complexity on the loss function and avoid over-fitting of the network~\cite{dl}. In addition, Table I lists the hyperparameters in the simulation results.

Fig. \ref{fig:reward} (a) examines the average reward per episode of the EMRQN and DQN algorithms. We set the discount factor as $\gamma=0.999$ to give a high weight for future reward~\cite{RL}. First, we find that the reward goes up as timestep increases and reaches the maximum average reward when the learning process becomes stable. Second, EMRQN converges to a larger average reward than DQN. Third, in view of the magnified areas, EMRQN has smaller fluctuation range than DQN. This is due to the fact that LSTM is more suitable for sequential decision-making problem than common NN and RNN. Fig. \ref{fig:reward} (b) shows different average step rewards of EMRQN under $\gamma$ = 0.999, 0.99, 0.90, and 0.85 respectively. First, we find that the step reward goes up as timestep increases and reaches the peak value when the learning process becomes stable. Second, the higher $\gamma$, the slower the convergence becomes. Since the training is processed offline, the time for training is not a major concern in this paper. This is due to the fact that the BS pays more attention to the future rather than the present. Third, the larger $\gamma$ also contributes to the larger peak value, which is beneficial for the BS to make the long-term decision towards higher cache hit rate.
\begin{table}[t]\label{table}
\centering	
\caption{Algorithm Hyperparameters}
\setlength{\tabcolsep}{1mm}{
\begin{tabular}{p{3.5cm}|p{2.2cm}|p{2.2cm}}  
\hline
\hline
\textbf{Parameters} & \textbf{EMRQN} & \textbf{DQN}\\
\hline
Learning rate & 0.00015 & 0.0002\\
\hline
Experience replay size & 100000 & 100000 \\
\hline
Optimizer & Adam & Adam\\
\hline
Initializer & Kaiming & Kaiming\\
\hline
Loss function & Huber loss & Huber loss\\
\hline
\hline
\end{tabular}}
\end{table}

Fig. \ref{fig:hit} compares the average cache hit rates among LRU, FIFO, DQN, and the proposed EMRQN algorithm with Zipf parameters of 1.5 and 0.8 respectively. We also consider a simple strategy that evicts the least requested content. Consider that the BS serves 20 users and the cache size varies from 250 to 1500. First, the cache hit rate goes up with increase of cache size. Second, EMRQN significantly outperforms the other four algorithms, and the simple strategy yields the worst cache hit rate. Third, cache hit rates of all algorithms are reduced when the Zipf parameter becomes smaller.
\section{Conclusion}\label{sec:end}
In this letter, we have developed a novel content update strategy to improve cache hit rate in the BS. Meanwhile, we have formulated the content update process as an MDP and put forth the EMRQN algorithm to enhance the decision making ability of the BS. Compared with conventional cache replacement algorithms, the proposed algorithm has gained a significant improvement in cache hit rate and long-term system reward. This work only considered the content update problem of a single BS. In practice, it is of interest to investigate a general caching network where multiple BSs cooperatively serve the users by updating their local caches. Due to the mutual effects on decisions among the BSs, how to share their caching states with minimal overhead remains challenging.
\bibliographystyle{IEEEtran}
\bibliography{cache}
\end{document}